\documentclass[letter]{aa}  

\usepackage{graphicx}
\usepackage{txfonts}
\usepackage{lipsum}
\usepackage{subcaption}         
\usepackage{lscape}             
\usepackage{placeins}           
\newcommand{\mum}{\textmu{}m\xspace}

\newcommand{\classA}{$\mathcal{A}$\xspace}
\newcommand{\classB}{$\mathcal{B}$\xspace}

\begin{document}

   \title{JWST MIRI-MRS observations of the Red Rectangle:\\ AIB class transformation in the outer nebula}

%

   \author{A. Ebenbichler\inst{1,2}\fnmsep\thanks{Corresponding author: aebenbic@uwo.ca}
        \and P. Moraga Baez\inst{1,2}
        \and A. Candian\inst{3}
        \and E. Peeters\inst{1,2}
        \and J. Cami\inst{1,2}
        \and P.J. Sarre\inst{4}
        \and A.N. Witt\inst{5}    
        \and I. Argyriou\inst{6}
        \and B. Vandenbussche\inst{6}
        \and H. Van Winckel\inst{6}
        \and L.B.F.M. Waters\inst{7,8}
        }

   \institute{Department of Physics \& Astronomy, The University of Western Ontario, London, ON N6A 3K7, Canada
   \and Institute for Earth and Space Exploration, The University of Western Ontario, London, ON N6A 3K7, Canada
   \and Anton Pannekoek Institute for Astronomy, University of Amsterdam, Science Park 904, 1098 XH, Amsterdam, The Netherlands
   \and School of Chemistry, University of Nottingham, University Park, Nottingham, NG7 2RD, United Kingdom
   \and Ritter Astrophysical Research Center, University of Toledo, Toledo, OH 43606, USA
   \and Institute of Astronomy, KU Leuven, Celestijnenlaan 200D, 3001 Leuven, Belgium
   \and SRON Netherlands Institute for Space Research, Niels Bohrweg 4, 2333 CA Leiden, The Netherlands
   \and Institute for Mathematics, Astrophysics and Particle Physics, Radboud University, 
   MC 62 NL-6500 GL Nijmegen, The Netherlands
   }

 
  \abstract
   {}
   {We characterize the mid-infrared spectrum of the outer regions of the Red Rectangle nebula to probe the carbonaceous dust and molecular content beyond the circumbinary disk. 
   }
   {We present JWST MIRI-MRS observations of the SW whisker, extracted from three distinct environments: the biconical outflow, the whisker itself, and the shadow region outside the outflow. We compare these with an archival ISO-SWS observation of the inner nebula. 
   }
   {The JWST spectra display only classical AIB emission on a weak dust continuum, with no signatures of the oxygen-rich circumbinary disk mineralogy nor of the rich molecular emission seen at optical wavelengths. 
   The AIBs are predominantly Class \classA -- in marked contrast to the exclusively Class \classB profiles previously reported for the inner regions -- with systematic differences between the outflow and shadow regions pointing to environmentally driven PAH processing. }
   {}


   \maketitle
\nolinenumbers
\section{Introduction}\label{sec:intro}
The Red Rectangle (RR) is one of the few mass-losing carbon-rich post-AGB objects with a spatially resolved 
nebula, associated with the central star HD~44179 \citep{1975ApJ...196..179C}. 
At the core is a binary system, in which the evolved primary ($M\approx0.8\,$M$_\sun$) feeds an accretion disk around a solar-mass main sequence companion. Temperatures in the inner circumcompanion disk reach  $\sim$17\,000\,K, sufficient to produce a compact H\,{\sc ii} region and to supply UV photons ($E<13.6$\,eV) to all parts of the nebula with direct lines of sight to the disk \citep{2009witt, 2023bujarrabal}. 
The HD~44179 binary system is also embedded in an optically thick, dusty circumbinary disk \citep{1995A&A...293L..25V,2013A&A...557L..11B,2016A&A...593A..92B}, seen nearly edge-on. This geometry, combined with the relative proximity of the RR \citep[$\sim710$\,pc;][]{2002A&A...393..867M, 2004cohen}, 
makes it possible to spatially resolve carbonaceous condensation and processing products across distinct environments: the UV-bathed outflow cones, the shadowed regions outside the cones, and the interfaces between the two that show up as the bright whiskers (the "X"-shape).
Each of these environments hosts distinct optical emission phenomena. The Extended Red Emission (ERE), responsible for the nebula's characteristic color, peaks at the interface between the outflow and shadow regions \citep{1991ApJ...383..698S, 1992MNRAS.255P..11S,2006ApJ...653.1336V}.
Superimposed on the ERE, the RR spectrum exhibits a unique set of narrow emission bands - the Red Rectangle Bands (RRBs) - at wavelengths closely matching those of some of the stronger diffuse interstellar bands  \citep[DIBs; ][]{1995Sci...269..674S,2002VanWinckel, 2011A&A...533A..28W}.
In contrast, Blue Luminescence (BL), peaking at 380\,nm, dominates the regions shadowed by the circumbinary disk.
The circumbinary disk contains large grains \citep{2023bujarrabal} and characteristically oxygen-rich material, including crystalline silicates and CO$_2$ gas, likely ejected from the outer envelope during an earlier evolutionary stage \citep{1998Natur.391..868W}.
Imaging surveys show that the infrared (IR) continuum is dominated by the central source \citep{2004A&A...415..179M, 2011MNRAS.417...32L}, while the outer regions trace the "X"-shaped morphology seen in the optical.
The IR spectrum is dominated by aromatic infrared bands (AIBs), particularly in the biconical outflow \citep{1998Natur.391..868W, 2007song, 2012MNRAS.426..389C}, and exhibits exclusively 
Class \classB profiles \citep{2002A&A...390.1089P} both in ISO observations of the full nebula  \citep{1998Natur.391..868W} and in ground-based sub-arcsecond spectroscopy of the innermost outflow regions 
\citep{2004A&A...415..179M}.
Crucially, all previous IR spectroscopic surveys have been limited to the central region of the nebula.
JWST now makes it possible to observe the outer regions with unprecedented sensitivity and spatial detail.
Here, we present JWST MIRI-MRS data covering the south-west whisker of the RR. 

\begin{figure}
    \centering
\includegraphics[width=0.99\linewidth, trim={3.4cm 0.7cm 3.5cm 0.5cm}, clip]{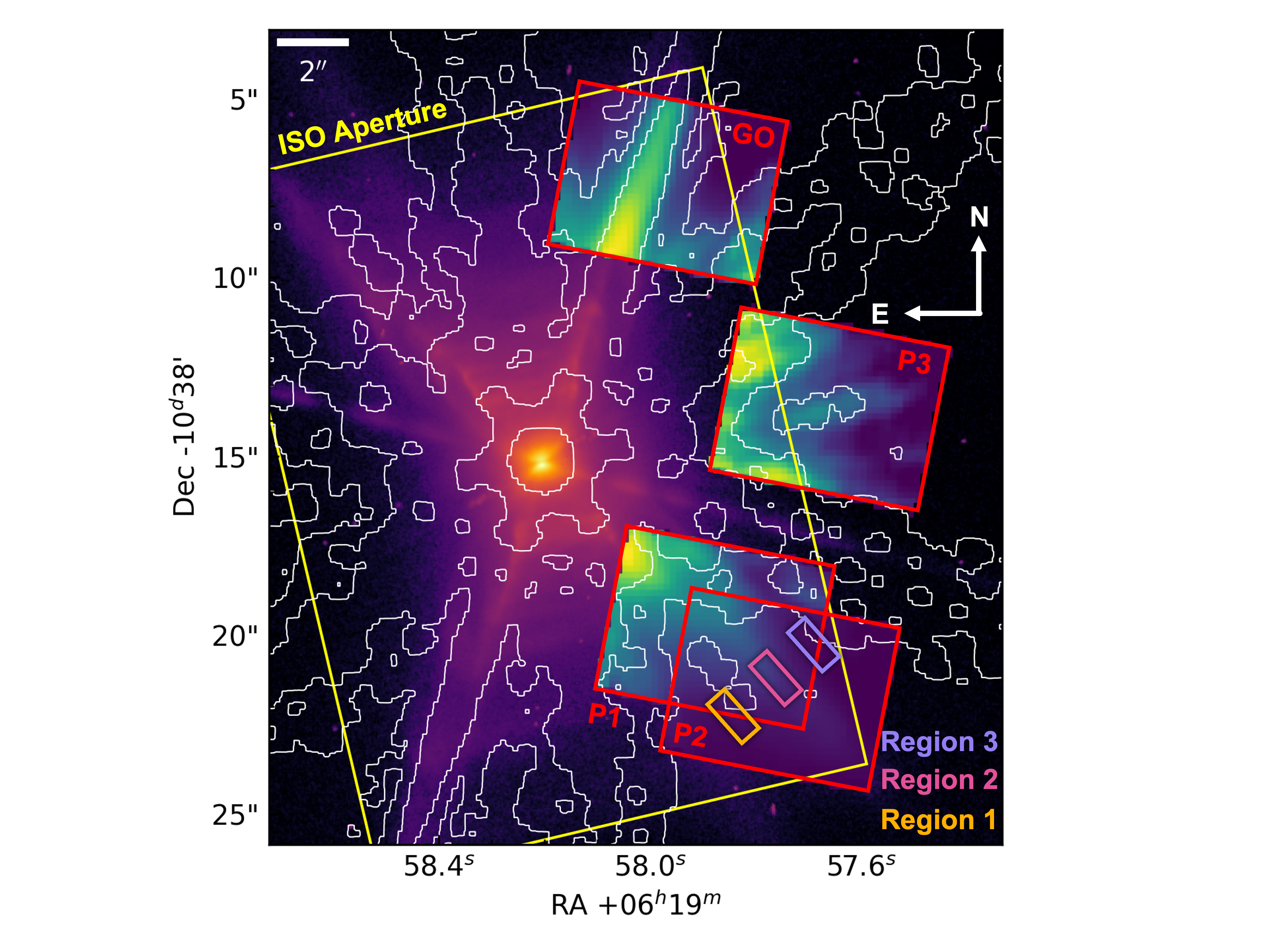}
    \caption{JWST-MIRI channel 2C apertures (summed from 10.01 to $11.7\,\mu$m) of the positions GO (north), P1P2 (south, mosaic of P1 and P2) and P3 (west) outlined in red on an HST WFPC2 F622W image. White contours represent a diffraction model of the central region at $10\,\mu$m, using the MIRI imaging point spread function (PSF).
    }
    \label{fig:FOVs}
\end{figure}

\section{Observations}\label{sec:obs}
JWST MIRI Medium Resolution Spectroscopy (MRS) Integral Field Unit \citep[IFU,][]{MIRI, MRS, MIRIperformance, MRSperformance}  observations of HD~44179 were obtained on 2025-02-18 (PID 5749, 4 point extended source dither; PI: A.~Candian) at four positions (Fig.~\ref{fig:FOVs}).
We used the level 3 pipeline products from the MAST JWST archive for positions GO and P3 (Calibration Software version 1.19.1, CRDS reference file context 1413). 
For the mosaic P1P2, consisting of the positions P1 and P2, we downloaded the level 2 products from the MAST JWST archive and produced the mosaic using the default JWST pipeline\footnote{\url{https://jwst-pipeline.readthedocs.io/en/latest/}} (version 1.20.2) and the CRDS reference file context 1464.

The central IR source is very bright, producing prominent diffraction spikes that affect all four MIRI positions (Fig.~\ref{fig:FOVs}). Unfortunately, the limited JWST visibility windows of the RR severely constrain the available (V3) position angles, leaving little room to orient the diffraction spikes away from the whiskers.
We performed PSF modeling to quantitatively assess the influence of diffraction at the different observed locations.
Position GO is the most severely affected, with a spike from the secondary mirror support strut coinciding with the NW whisker. 
For position P3, diffraction from the central source dominates the field entirely, precluding straightforward analysis. 
For the P1P2 mosaic however, the nebular surface brightness exceeds that of the central source diffraction over most of the field. 
We therefore concentrate on this mosaic for our analysis, using stitched cubes as final data products. 
Further details on the central source diffraction are given in Appendix~\ref{sec:psf}.

\section{Results}\label{sec:res}
\begin{figure*}
    \centering
\resizebox{\hsize}{!}{
    \includegraphics[clip, trim=0.3cm .45cm 0.4cm .3cm]{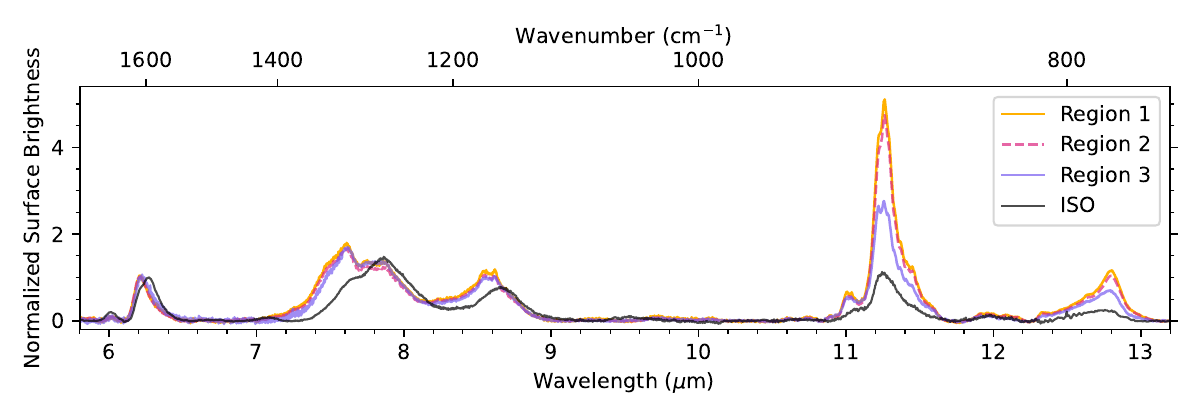}}
    \caption{JWST/MIRI spectra from Regions 1--3 compared with the ISO-SWS spectrum (black). All spectra are continuum-subtracted and normalized to the peak of the 6.2 $\mu$m AIB band. Apertures are defined in Fig.~\ref{fig:FOVs}.}
    \label{fig:Aps_full}
    
\end{figure*}

\begin{figure}
    \centering
    \resizebox{.88\hsize}{!}{
    \includegraphics[trim={0cm 0.5cm 0cm 0cm}, clip]{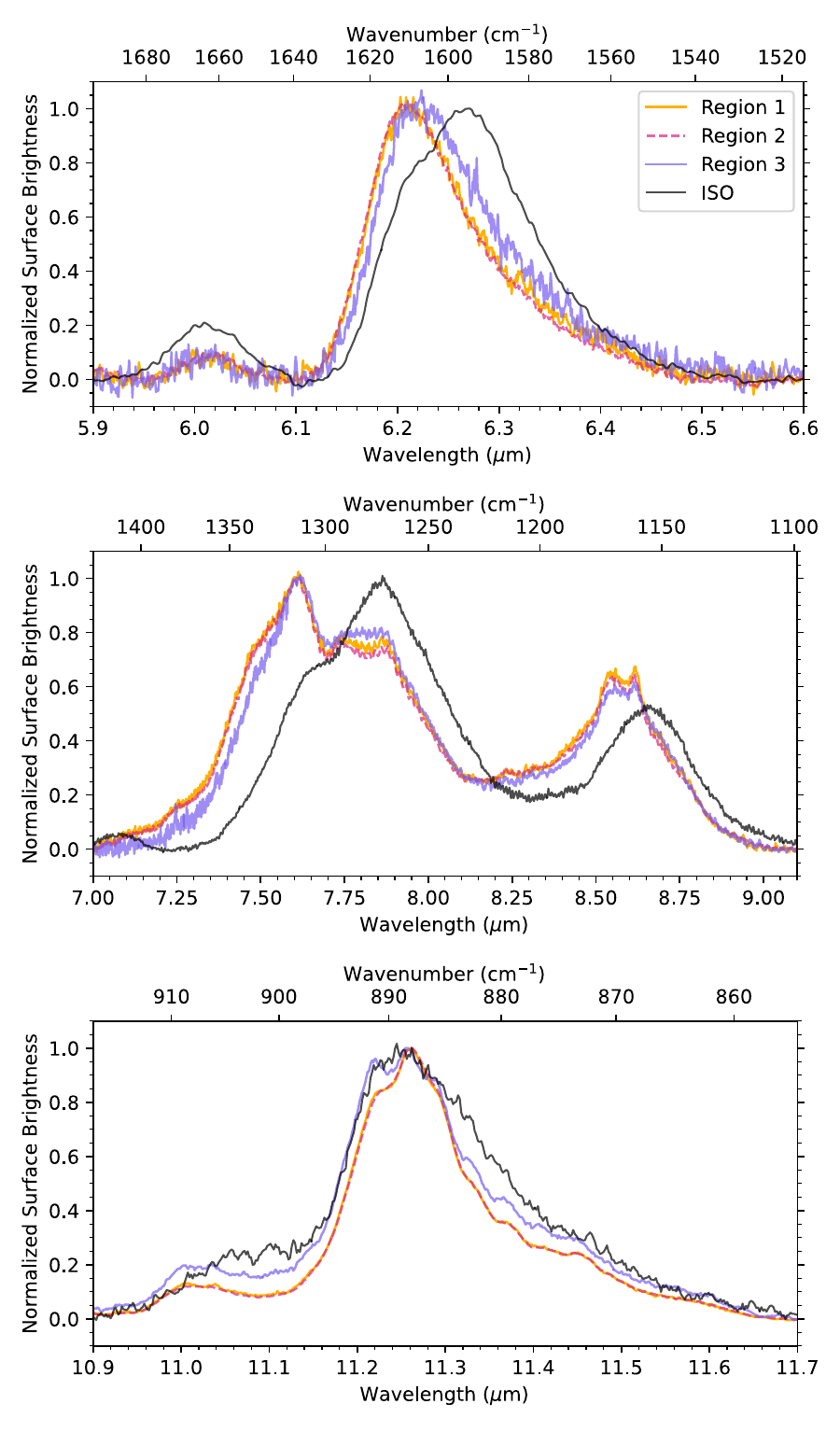}}
    \caption{AIB profile comparison for the 6.2 $\mu$m (top); 7.7 $\mu$m and 8.6 $\mu$m (middle); 11.2 $\mu$m (bottom) AIB features. 
    Spectra are normalized to the dominant AIB in each panel. }
    \label{fig:Aps}
\end{figure}

Figs.~\ref{fig:Aps_full}, \ref{fig:Aps} and \ref{fig:Aps_full_cont} show JWST-MIRI P1P2 spectra (Channels 1A--3A), extracted from three regions: within the biconical outflow (Region 1), on the SW whisker (Region 2), and outside the biconical outflow (Region 3). For comparison, we include the archival ISO-SWS spectrum\footnote{AOT 1 speed 4, TDT (Target Dedicated Time) number 70201801}, whose $14'' \times 20''$ aperture is centered on the strongly emitting circumbinary disk \citep[see Fig.~\ref{fig:FOVs}; ][]{1998Natur.391..868W,2002A&A...390.1089P}. Notably, all three JWST-MIRI spectra exhibit only AIB emission superimposed on a very weak dust continuum. 
Comparison of the JWST and ISO spectra reveals pronounced differences in the relative band ratios (Fig.~\ref{fig:Aps_full}).
In particular, the 11.2 and 12.7\,\mum AIBs are significantly enhanced relative to the 6--9\,\mum AIBs in the JWST observations compared to the ISO spectrum, with this enhancement being most pronounced in Regions 1 and 2  and weaker in Region 3. 
In contrast, the 12.0/6.2~$\mu$m AIB ratio remains roughly constant across all observations. 

Turning to the AIB profile shapes, the JWST spectra overall exhibit Class \classA AIB characteristics which are commonly found in the interstellar medium, including {\sc H\,ii} regions and reflection nebulae. This contrasts sharply with the Class \classB spectrum previously observed with ISO-SWS (Fig.~\ref{fig:Aps}), which is more common in circumstellar environments \citep{2008ARA&A..46..289T}. This indicates that AIB carriers in the circumbinary disk and in the 
outflow have undergone different processing, resulting in distinct class types. 

\textbf{The 6.2 \mum feature:} The spectra from Regions 1 and 2 show Class \classA characteristics, most notably a peak at 6.21~$\mu$m.  
The Region 3 spectrum also classifies as Class \classA, but its peak is redshifted by $\sim0.01$\,$\mu$m relative to Regions 1 and 2, despite the sharp blue rise having the same onset wavelength. The red wing is also more pronounced in Region 3. In contrast, the ISO profile peaks at 6.268~\mum \citep{2002A&A...390.1089P}, is considerably broader, and contains a component at approximately 6.205~\mum corresponding to the peak position of the JWST spectra.

\textbf{The 7--9 $\mu$m complex: }
The JWST spectra display class \classA profiles for the 7.7~\mum complex, with a dominant 7.6~\mum component and a weaker 7.8~\mum component. The 7.6~\mum component features a relatively narrow peak with a clear blue shoulder at 7.45~\mum and a weaker subcomponent at 7.25~\mum. In Region 3, the blue wing is redshifted relative to Regions 1 and 2, with a less prominent shoulder at 7.50~\mum, and the 7.7--7.9~\mum range shows higher relative intensity. The spectra of Regions 1 and 2 reveal double-peaked profiles in this range, with peaks at 7.705 and 7.851~\mum, similar to those recently observed in the Orion Bar \citep{2024A&A...685A..75C}. Interestingly, the second peak coincides with the ISO 7.8 $\mu$m feature position. The 8.6 $\mu$m feature also shows a double-peaked profile (at 8.54 and 8.62~\mum) with weak shoulders at 8.45 and 8.70~\mum, differing markedly from the smoother ISO profile. 
Overall, the entire 7--9~\mum AIB complex in the JWST observations is shifted to shorter wavelengths relative to the ISO observations, consistent with the Class \classA designation \citep{2002A&A...390.1089P}. Additionally, a 7.05~\mum aliphatic feature present in the ISO spectrum is absent in the JWST spectra. 

\textbf{The 11.2 $\mu$m feature: }
The ISO spectrum shows a Class \classB 11.2\,$\mu$m AIB. The JWST spectra also classify as Class \classB based on peak position, but exhibit a blue shoulder at 11.21~$\mu$m that falls within the Class \classA peak position range \citep{2004diedenhoven}. Also, the JWST profiles have a considerably less pronounced red wing and thus a narrower profile than is typical for Class \classB \citep{2004diedenhoven}: the FWHM of the 11.2~$\mu$m AIB in Regions 1 and 2 is comparable to that of Class \classA profiles \citep{2004diedenhoven}. 

\section{Discussion}\label{sec:disc}
The JWST-MIRI data of the outer nebula display only AIB features on a weak continuum, with no signatures of the oxygen-rich mineralogy that dominates the central regions. Previous spectroscopy \citep[see e.g.][]{1993ApJ...409..412S, 2004A&A...415..179M} established that the circumbinary disk itself shows no AIB features -- with the possible exception of the $3.3\,\mu$m band, though this detection was obtained with a beam size of 0.6--0.7$^{\prime\prime}$ \citep{2003MNRAS.346L...1S, 2012MNRAS.426..389C}, considerably larger than the $0.3\,''$ PSF of \citet{2004A&A...415..179M}.
Instead, the disk spectrum is characterized by crystalline olivine features, CO emission and CO$_2$ absorption, pointing to an oxygen rich chemistry \citep{1998Natur.391..868W}.
At the central source position, the only feature identified by \citet{2004A&A...415..179M} at longer wavelengths is at $12.0\,\mu$m, likely due to crystalline olivine \citep{1998Natur.391..868W}. Dust emission is confined to the circumbinary disk, while AIBs dominate the IR emission in the biconical outflow \citep{2009gledhill}.

In the JWST observations of the outflow, the AIBs are Class \classA with the exception of the $11.2\,\mu$m feature -- in strong contrast to all previous studies of the inner regions, which found exclusively Class \classB \citep{1993ApJ...409..412S, 1998Natur.391..868W, 2002A&A...390.1089P, 2004A&A...415..179M}.
In previous ground-based data, the aromatic $3.3\,\mu$m feature is visible on-star and along the outflows, while the 
aliphatic $3.4\,\mu$m feature is undetectable on-star but increases relative to the $3.3\,\mu$m feature with distance from the center, reaching a plateau at $4\,''$ \citep{2007song}.
The $3.30\,\mu$m peak shifts to the blue with increasing distance.
In the N band, \citet{2004A&A...415..179M} describe the spectral evolution from 0 to $1.7\,''$ (PSF FWHM $0.3\,''$): an $8.6\,\mu$m feature and an 11.2\mum feature blended with a strong $11.0\,\mu$m component appear at 0.3'', after which the $11.0\,\mu$m feature weakens and narrows relative to the $11.2\,\mu$m band.
In our data, the $11.2\,\mu$m profile narrows further along the outflow, consistent with a decrease in vibrational temperature due to processing and destruction of less stable PAHs, a decreasing radiation field, or the destruction of very small grains and PAH clusters \citep{2024A&A...685A..75C, 2024A&A...685A..77P, khan_pdrs4all_2025}.

The hypothesis of progressive PAH processing is supported by the overall Class \classB-to-\classA transformation: the 6--9\mum features all shift to shorter wavelengths between the ISO (inner) and JWST (outer) observations, which is generally attributed to more processed PAH populations. \citet{2004A&A...415..179M} already observe the $8.6\,\mu$m peak shifting from 8.70 to $8.65\,\mu$m between 0.7 and $1.7\,''$, reflecting changes within Class \classB towards Class \classA. This trend continues at $9\,''$ in our Regions 1 and 2, where the $7.7\,\mu$m complex peaks at $7.61\,\mu$m. However, a few studies argue that the 7.9\,\mum component \citep{Joblin2008} or Class \classB \citep{2025smithperez} is due to more processed PAH populations.
The $11.2\,\mu$m feature does not follow the same trend. Its peak shifts from 11.27 to $11.23\,\mu$m between 0.7 and $1.7\,''$ \citep{2004A&A...415..179M} but settles at $11.26\,\mu$m at $9\,''$ -- between the values in the center -- showing no clear monotonic evolution.
This reinforces the suggestion by \citet{2025smithperez} that the Class \classB 11.2\mum profile as described by \citet{2004diedenhoven} might not exist as a distinct category.

The complete transformation from Class \classB in the center to Class \classA at $9''$ is notable, as PAH processing is generally driven by intense UV radiation.
Compared to similar objects, the RR's accretion disk ($T_\mathrm{eff} = 17\,000\,$K) produces a high flux of UV photons with energies $E > 10.5\,$eV, sustained by a mass accretion rate of $\sim 2 \times 10^{-5}\,$M$_\sun \mathrm{yr}^{-1}$ \citep{2009witt} -- sufficient to photo-process ERE carrier precursors \citep{2006witt}, and, presumably, PAHs as well.
The absence of H$_2$ emission indicates that the outflow region is an atomic PDR and limits the radiation field to $G_0/n < 0.01 - 0.1\,\mathrm{cm}^{3}$ \citep[$G_0$ is far-UV intensity in Habing units, $n$ is total number density;][]{1997ARA&A..35..179H}
Shocks represent a second possible processing mechanism, shown to be viable in theoretical studies \citep{2010micelotta}.

In addition to the spectral evolution with distance from the center, we observe systematic differences between the biconical outflow (Regions 1 and 2) and the shadow region (Region 3).
The spectra of Regions 1 and 2 are very similar (Fig.~\ref{fig:Aps}),
as expected given that the two apertures lie at comparable distance from the source and probe similar conditions along the symmetric outflow. Region 2 shows stronger overall emission, as its line of sight intersects the cone tangentially, but the band ratios and profiles closely match those of Region 1.
Region 3, however, shows significant differences.
The redshift of the $6.2\,\mu$m peak, the shift of the $7.7\,\mu$m complex to longer wavelengths, and the changes in the $11.2\,\mu$m profile all indicate a slight shift towards Class \classB in the shadow region. This is consistent with reduced UV irradiation and less dynamic processing outside the outflow, leading to different excitation conditions and chemical structure of the PAH population.
The 6.2/11.2 band ratio increases markedly from Region 2 to Region 3. While this might suggest an increase in PAH ionization, the weaker radiation field in the shadow region makes this unlikely.
An alternative explanation is a reduction in the number of solo-H sites on neutral PAHs, possibly through super-hydrogenation. In this scenario, the PAH mass distribution would remain largely unaffected, but $11.2\,\mu$m emission would be suppressed by this chemical modification. We note that the 6.2/12.7 ratio is also reduced, though to a lesser extent. A second possibility is that the variations in the C-H out-of-plane bending modes (10-15~\mum region) reflect differences in the PAH size distribution: smaller PAHs may be preferentially destroyed within the outflow, increasing the fraction of large PAHs (and so the $11.2\,\mu$m), while they survive more readily in the sheltered shadow region.
Distinguishing between these scenarios requires the $3.3/11.2\,\mu$m ratio, which best traces the PAH size distribution, but NIR observations of this location on the SW whisker are currently lacking.

\section{Conclusions}\label{sec:conc}
JWST MIRI-MRS observations of the SW whisker of the Red Rectangle reveal a spectrum dominated entirely by AIB emission on a weak dust continuum, with no trace of the oxygen-rich mineralogy confined to the circumbinary disk. Most AIB features in the outer nebula are Class \classA, in marked contrast to the Class \classB profiles found in all previous observations of the inner regions — with the notable exception of the 11.2~\mum band, whose anomalous behavior supports the suggestion by \citet{2025smithperez} that the classical Class \classB designation for this feature may need revision. The Class \classB-to-\classA transformation along the outflow is consistent with progressive PAH processing, plausibly driven by the strong UV field from the accretion disk or by shocks. Systematic differences between the outflow and shadow regions — including shifts towards Class \classB and an elevated 6.2/11.2 ratio outside the outflow — point to the role of the local environment in shaping PAH chemistry, with super-hydrogenation or differential size-dependent survival as possible explanations. Remarkably, despite the rich optical phenomenology of the RR (ERE, RRBs, Blue Luminescence), the mid-infrared spectrum shows only classical AIB features. More detailed analyses of these data, including spatially resolved mapping and spectral decomposition, will be presented in forthcoming studies (Moraga Baez et al., in prep.; Ebenbichler et al., in prep.).

\begin{acknowledgements}
This work is based on observations made with the NASA/ESA/CSA James Webb Space Telescope. The data were obtained from the Mikulski Archive for Space Telescopes at the Space Telescope Science Institute, which is operated by the Association of Universities for Research in Astronomy, Inc., under NASA contract NAS 5-03127 for JWST. These observations are associated with program \#5749 (DOI:10.17909/fe3t-2e17).
Astrochemistry in the Netherlands is supported by the Dutch Astrochemistry Network of the Dutch Research Council (NWO) under grant no. ASTRO.JWST.001.
E. Peeters and J. Cami acknowledge support from the Canadian Space Agency (CSA, 24JWGO3A14), and the Natural Sciences and Engineering Research Council of Canada. 
This article is based upon work from COST Action CA21126 - Carbon molecular nanostructures in space (NanoSpace), supported by COST (European Cooperation in Science and Technology). 
\end{acknowledgements}

\bibliographystyle{aa} 
\bibliography{bibliography} 
\listofobjects

\begin{appendix}
\section{Diffraction effects from the central infrared source}\label{sec:psf}

The detailed nature of the source of the strong diffraction spikes is not trivial, and has to be discussed thoroughly, as it poses a major contamination of our science data.
Most of the RR's flux density continuum is contributed by the circumbinary disk around HD~44179 \citep{2002A&A...393..867M}. 
The densest part of this disk is rather compact with a radius of $r\approx 0.09\,''$ in the mid-infrared \citep{2004A&A...415..179M}.
However, multiple studies have shown that the disk mainly emits continuum radiation but no emission bands. 
Using long slit observation between 7 and 14\,$\mu$m, \citet{1993ApJ...409..412S} showed that the disks contribution can be approximated by a black body curve with a temperature of $B_\nu=370$\,K and is featureless in this wavelength range. 
The AIBs emerge in the bipolar outflow, with the $7.7\,\mu$m and $8.6\,\mu$m features at $0.4 - 0.9\,''$, and the $11.2\,\mu$m feature at $0.9 - 1.8\,''$ from the disk. 
\citet{2004A&A...415..179M} found a similar trend of a relatively featureless spectrum in the centre, however with a feature at $12\,\mu$m.
Given that some AIBs are very strong close to the centre, it is possible that they get diffracted onto our JWST-MIRI observations.
As we do not know if all AIBs are absent in the diffraction pattern, we used the ISO-SWS spectrum to estimate the diffracted spectrum as an upper limit. 
We multiplied the ISO SED with PSF models using the Python package STPSF \citep{2014SPIE.9143E..3XP}. 
Then, we scaled the surface brightness by fitting the diffraction model to channel 4 in P3, which is strongly dominated by the diffraction.
We also noticed a spatial offset between the diffraction pattern in our models and the observations.
This is a result of a small pointing error of JWST of $RA_\mathrm{GAIA} - RA_\mathrm{JWST} \equiv \Delta_{RA} = 0.23\pm0.05\,''$ and $DEC_\mathrm{GAIA} - DEC_\mathrm{JWST} \equiv \Delta_{DEC} = 0.23\pm0.07\,''$.
The pointing error was the same for all tested channels and positions within standard errors, which are smaller than the MIRI-MRS pixel scale.

As the ISO SED is an approximation, and STPSF can not model the spectral effects of the PSF, the resulting PSF data cubes can not be used for subtraction from the science data.
However, the PSF model gives a good approximation for the upper limit of the contribution of diffraction relative to the science data.
Using our PSF model, we determined that only the east and north-north west corners of the P1P2 mosaic are significantly affected, which are not used for our analysis.
We are working on refining our PSF model and will show the resulting progress in following publications (Moraga Baez et al., in prep.).

\section{Spectra with continuum}
\begin{figure*}
    \centering
    \subfloat{
    \includegraphics[width=0.89\linewidth, clip, trim=0cm .45cm 0cm .3cm]{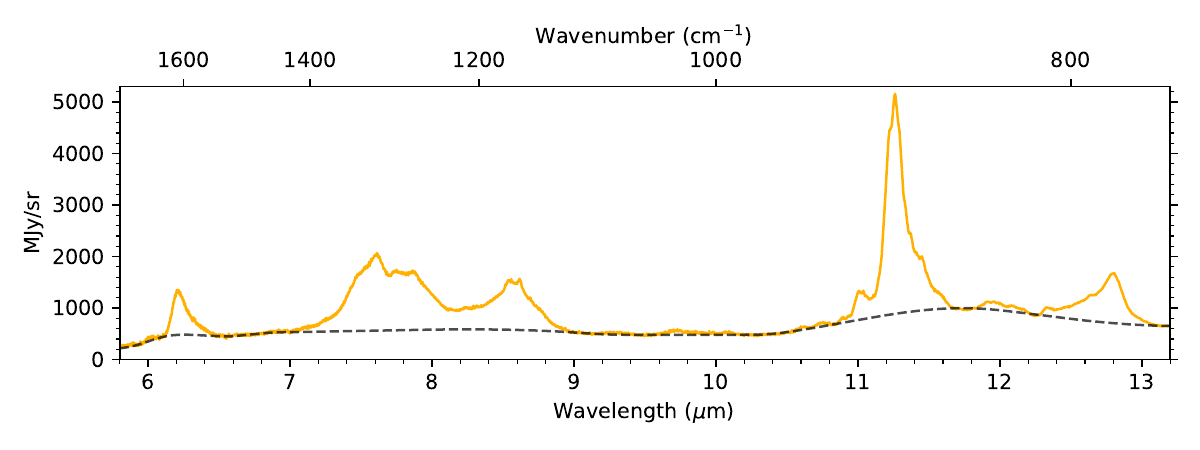}
    }\\
    \subfloat{
    \includegraphics[width=0.89\linewidth, clip, trim=0cm .45cm 0cm .3cm]{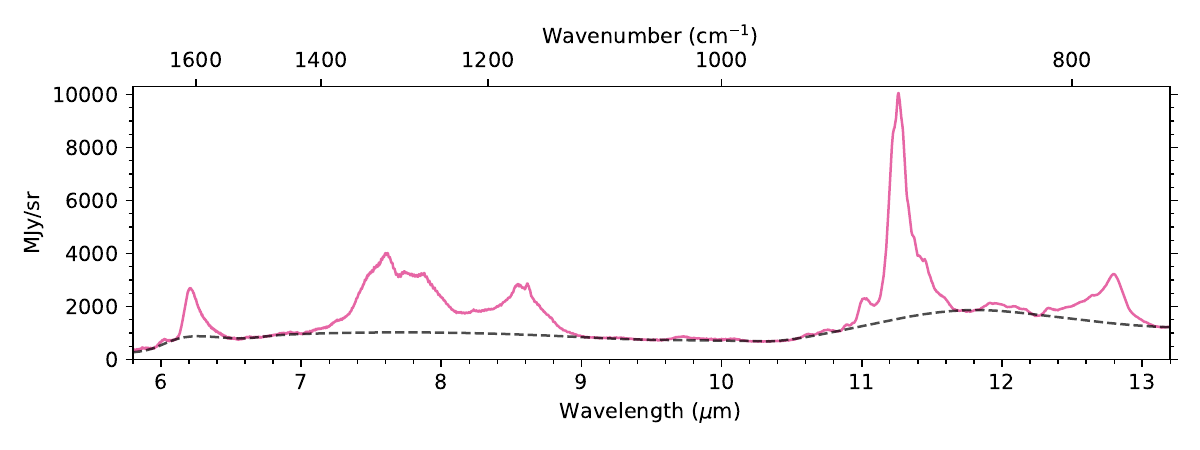}
    }\\
    \subfloat{
    \includegraphics[width=0.89\linewidth, clip, trim=0cm .45cm 0cm .3cm]{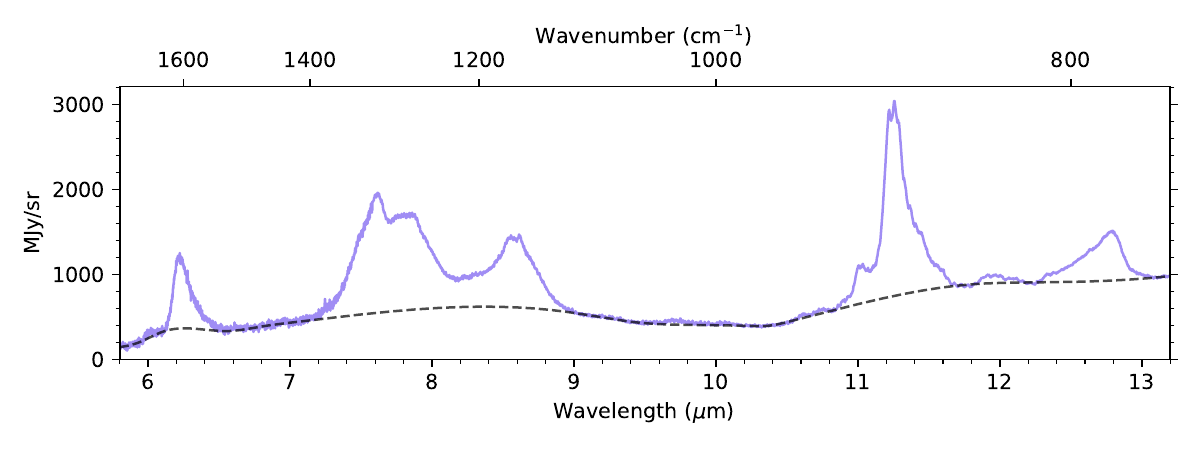}
    }\\
    \subfloat{
    \includegraphics[width=0.89\linewidth, clip, trim=0cm .45cm 0cm .3cm]{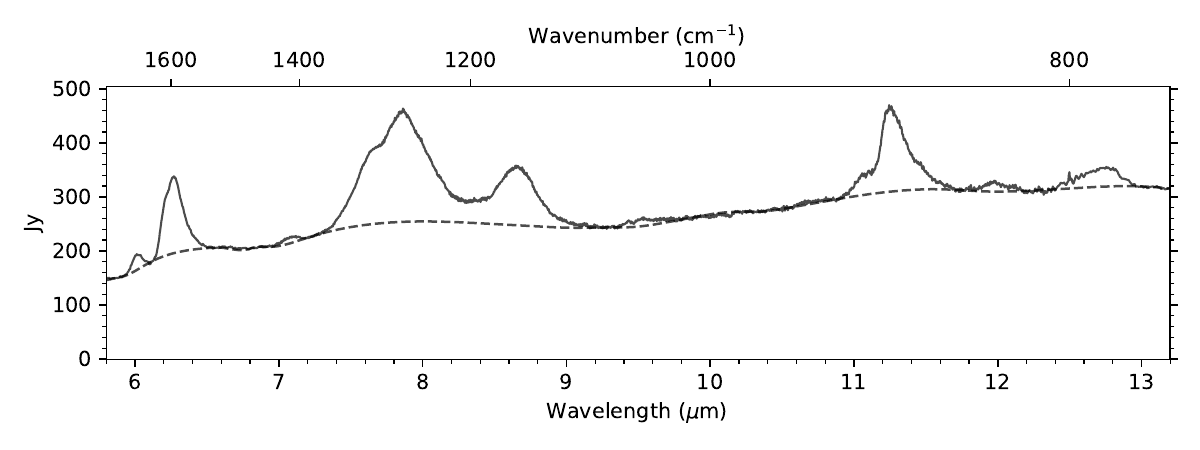}
    }\\
    \caption{Comparison of JWST-MIRI spectra (from top to bottom: Region 1, 2, and 3) with the ISO-SWS spectrum (bottom). The MIRI apertures are defined in Fig.~\ref{fig:FOVs}.
    The black dashed line indicates the continuum which was subtracted in Figs.~\ref{fig:Aps_full} and \ref{fig:Aps}. Note that the JWST spectra are in surface brightness units (MJy/sr) while the ISO-SWS spectrum is in flux density (Jy). }
    \label{fig:Aps_full_cont} 
\end{figure*}

In Fig.~\ref{fig:Aps_full_cont}, we show the original spectra of our Regions 1--3 and the ISO spectrum before continuum subtraction and normalization. The continuum contribution is significantly lower in the JWST data than in the ISO data.

\end{appendix}

\end{document}